\documentclass[12pt]{article}
\usepackage{amssymb,amsmath}
\begin{document}

\sloppy
\title
{\Large Gravity according to theory of sources }

\author
 {
       A.I.Nikishov
          \thanks
             {E-mail: nikishov@lpi.ru}
  \\
               {\small \phantom{uuu}}
  \\
           {\it {\small} I.E.Tamm Department of Theoretical Physics,}
  \\
               {\it {\small} P.N.Lebedev Physical Institute, Moscow, Russia}
  \\
 }
%
\maketitle
 \begin{abstract}
The metric of spherically symmetric ball of ideal liquid is considered in $G^2$- approximation with the help of theory of sources. Using the integral equations of this theory gives the exterior metric depending upon the radius of the ball of matter in some terms proportional to $G^2$..I argue that according this metric from measurement outside the ball one can
infer the value of ball radius. 
\end{abstract}

{\bf Interior solution}
The following notation is used
$$
 g_{ik}=\eta_{ik}+h^{(1)}_{ik}+ h^{(2)}_{ik}+\cdots,\quad h^{(n)}_{ik}\propto
G^n,\quad\eta_{ik}={\rm diag}(-1,1,1,1),\quad \frac{\partial h}{\partial x^{\alpha}}=h_{,\alpha},
$$
$$
\bar h^{(n)}_{ik}= h^{(n)}_{ik}-\frac12\eta_{ik}h^{(n)},\quad i,k=0,1,2,3,\quad h^{(n)}= h^{(n)}_{,\alpha\alpha}- h^{(n)}_{00},\quad \alpha,\beta=1,2,3.                                                                                                                                    \eqno(1)
$$
In linear approximation in Hilbert gauge $\bar h^{(1)}_{\alpha\beta.\alpha}(m',r)=0$ we have
$$
h^{(1)}_{ik}(m',r)=-2\phi(m',r)\delta_{ik},\quad\bar h^{(1)}_{ik}=-4\phi(m',r)\delta_{0i}\delta_{0k}, \quad
\phi(m',r)=\frac{m'G}{2b}(\frac{r^2}{b^2}-3),\quad
m'=\frac43\pi b^3\mu.                                                                                                                                                                  \eqno(2)
$$
Here $\mu$ is the density of liquid. The function 
 $\bar h^{(2)}_{ik}(m',r)$  in this gauge satisfies the differential equation  
$$
\nabla^2\bar h^{(2)}_{ik}(m',r)=-16\pi G(T^{(1)}_{ik}+t_{ik}),                                                                                                                                         \eqno(3)
$$
This is the differential form of eq. (17.6) in Ch.3, \S17 in [2], see also [1].

Here $T^{(1)}_{ik}$ is the tensor of matter, $t_{ik}$-tensor due to  3-graviton interaction In this paper we assume that $t_{ik}$ is given by general relativity, see Ch. 7, \S 6 in [3].

 In functions proportional to $G^2$, such as $\bar h^{(2)}_{ik}(m',r)$, $ h^{(2)}_{ik}(m',r)$ and so on $m'$
indicate only that these functions are obtained directly from equations of perturbation theory. In terms proportional to  $G^2$ in the considered approximation it does not matter whether it stands there $m'$ or the dressed mass $m=m'(1+3mG/3)$.
In the region inside the ball we have [1]
$$
t_{00}=-\frac{3}{8\pi G}(\nabla\phi)^2-6\mu\phi=\frac{m^2G^2}{8\pi Gb^4}(54-\frac{21r^2}{b^2}),\quad
 T_{00}^{(1)}=2\mu\phi=\frac{m^2G^2}{4\pi Gb^4}(\frac{3r^2}{b^2}-9),                                                                                                                                     \eqno(4)
$$
$$
T_{00}^{(1)} +t_{00}=\frac{m^2G^2}{8\pi Gb^4}(36-\frac{15r^2}{b^2}), 
$$
$$
T_{\alpha\beta}^{(1)}+t_{\alpha\beta}=\frac{m^2G^2}{8\pi Gb^4}[\delta_{\alpha\beta}\left(-9+\frac{4r^2}{b^2}\right)-\frac{2x_{\alpha}x_{\beta}}{b^2}].
                                                                                                \eqno(5)
$$ 

Now we show that  $\bar h^{(2)}_{ik}$ is defined by the differential equation  (3) up to an additive constant $c_0$. From (3) and (5) it is seen that the general form of solution of (3) should be a polynomial of degree four in $r$, because the source (5) is the polynomial of degree two and the operator $\nabla^2$ decreases the degree by two. So
$$
\bar h^{(2)}_{ik}(m',r)=(\frac{mG}{b})^2[\delta_{\alpha\beta}(c_0+c_2\frac{r^2}{b^2}+c_4\frac{r^4}{b^4})+
c_5\frac{x_{\alpha}x_{\beta}}{b^2}+c_6\frac{r^2x_{\alpha}x_{\beta}}{b^4}].                     \eqno(6)
$$
Using this form in (3) and taking into account the relations 
$$
\nabla^2r^n=(n^2+n)r^{n-2},\quad \nabla^2(r^nx_{\alpha}x_{\beta})=2r^n\delta_{\alpha\beta}+
n(n+5)r^{n-2}x_{\alpha}x_{\beta}                                                                 \eqno(7)
$$
we find
$$
c_6=\frac27,\quad c_4=-\frac37,\quad 3c_2+c_5=9.                                               \eqno(8)
$$
Now we remember that our solution should satisfy Hilbert condition: $[\bar h^{(2)}_{\alpha\beta}]_{,\alpha}=0$.
Taking into account  the relations  
$$
(r^2)_{,\beta}=2x_{\beta},\quad (r^4)_{,\beta}=4r^2x_{\beta},\quad (x_{\alpha}x_{\beta})_{,\alpha}=4x_{\beta},\quad
 (r^2x_{\alpha}x_{\beta})_{,\alpha}=6r^2x_{\beta},                                                      \eqno(9)
$$
we find from Hilbert condition $c_2+2c_5=0$. From here with the help of (8) we get $c_2=\frac{18}{5}, c_5=-\frac95$.
So only  $c_0$ remains undefined. It can be obtained from the continuity condition with the exterior solution.

On the other hand using the integral equation of theory of sources, see [1]  and  Ch. 3, \S 17 in [2]
$$
\bar h_{ik}^{(2)}(m',r)=16\pi G\int d^4x'D_+(x-x')[T_{ik}^{(1)}(x')+t_{ik}(x')],                           \eqno(10)
$$
we find
$$
\bar h^{(2)}_{\alpha\beta}(m',r)=\frac{m^2G^2}{b^2}[\delta_{\alpha\beta}(-5+\frac{18}{5}\frac{r^2}{b^2}-
\frac{3}{7}\frac {r^4}{b^4})-
\frac{9}{5}\frac{x_{\alpha}x_{\beta}}{b^2}+\frac{2}{7}\frac {r^2x_{\alpha}x_{\beta}}{b^4}]= \bar h^{(2)}_{\alpha\beta}(m,r),                                                                                                         \eqno(11)       $$             
$$
\bar h_{00}^{(2)}(m',r)=\frac{m^2G^2}{b^2}(\frac{51}{2}-12\frac{r^2}{b^2}+\frac{3}{2}\frac{r^4}{b^4}).                                                                                                          \eqno(12)
  $$
 The expression (11) remains valid when  $m'$ is replaced by $ m$ because   $\bar h^{(1)}_{\alpha\beta}(m',r)=0$ in (2)  for
$\alpha,\beta=1,2,3$, see the text below eq. (16).
From (11) we have
$$
\bar h^{(2)}_{\alpha\alpha}(m',r)=\bar h^{(2)}_{\alpha\alpha}(m,r)=\frac{m^2G^2}{b^2}[-15+9\frac{r^2}{b^2}-\frac{r^4}{b^4}]. \eqno(11a)
$$

Next from solutions $\bar h_{ik}^{(2)}(m',r)$ we should obtain the solution $ h_{ik}^{(2)}(m',r)$:
$$
 h^{(2)}_{ik}(m',r)=\bar h^{(2)}_{ik}(m',r)-\frac{1}{2}\eta _{ik}\bar h(m',r),\quad \bar h^{(2)}=
=\bar h^{(2)}_{\alpha\alpha}-\bar h^{(2)}_{00}= -h^{(2)}.                                                                                                                                                          \eqno(13)
$$
Using the eqs. (11a) and (12), we get from the second eq. in (13)
$$
\bar h^{(2)}(m',r)=\frac{m^2G^2}{b^2}(-\frac{81}{2}+21\frac{r^2}{b^2}-\frac52\frac{r^4}{b^4}).                                                                                                                  \eqno(14)
$$                                     
From the first eq. in (13) (in which $i,k$ are replaced by $\alpha,\beta$) we obtain
$$
h^{(2)}_{\alpha\beta}(m',r)=\frac{m^2G^2}{b^2}[\delta_{\alpha\beta}(\frac{61}{4}-\frac{69}{10}\frac{r^2}{b^2}+
\frac{23}{28}\frac{r^4}{b^4})-
\frac95\frac{x_{\alpha}x_{\beta}}{b^2}+\frac{2}{7}\frac{r^2x_{\alpha}x_{\beta}}{b^4}].                                                                                                                                                                                                                                  \eqno(15)
$$
Similarly from  (13) when $i=k=0$ we find
$$
h^{(2)}_{00}(m',r)=\frac{m^2G^2}{b^2}(\frac{21}{4}-\frac32\frac{r^2}{b^2}+\frac14\frac{r^2}{b^2}). \eqno(16)
$$

Besides solutions $h^{(2)}_{ik}(m',r)$ we need solutions $h^{(2)}_{ik}(m,r)$. The latter solutions can be obtained from
the former ones by adding to them $G^2$-terms from $h^{(1)}_{ik}(m',r)$ when in them $m'$ is replaced by 
$m(1-3\frac{mG}{b})$; 
 $m$ is the dressed mass, see [1] or equation (4.34) in [4]. Indeed we have 
$$
h^{(1)}_{ik}(m',r)=\frac{m'G}{b}\delta_{ik}(3-\frac{r^2}{b^2})=h^{(1)}_{ik}(m,r)+\frac{m^2G^2}{b^2}
\delta_{ik}(3\frac{r^2}{b^2}-9),\quad m'=m(1-3\frac{mG}{b}).                                                                                                                                               \eqno(2a)
$$
Similarly from (2)
$$
\bar h_{00}^{(1)}(m',r)=-4\phi(m',r)=\bar h_{00}^{(2)}(m,r)+\frac{m^2G^2}{b^2}(6\frac{r^2}{b^2}-18).                                                                                                          \eqno(2b)
$$
Equations (13) remain valid also when  $m'$ is replaced  by $m$. So taking into account  (11а) and equations  (12а) below from the second equation in (13) we find 
$$
\bar h^{(2)}(m,r)=\frac{m^2G^2}{b^2}(-\frac{45}{2}+15\frac{r^2}{b^2}-\frac52\frac{r^4}{b^4}),                                                                                                             \eqno(17)
$$
Similarly from the first equation in (13) with the help of (11) and (17) we get                                     
$$
h^{(2)}_{\alpha\beta}(m,r)=\frac{m^2G^2}{b^2}[\delta_{\alpha\beta}(\frac{25}{4}-\frac{39}{10}\frac{r^2}{b^2}+
\frac{23}{28}\frac{r^4}{b^4})-
\frac95\frac{x_{\alpha}x_{\beta}}{b^2}+\frac{2}{7}\frac{r^2x_{\alpha}x_{\beta}}{b^4}],                                                                                                                                                                                                                              \eqno(18)
$$
In the same manner with the help of  (12а) and (17) we obtain
$$
h^{(2)}_{00}(m,r)=\frac{m^2G^2}{b^2}(-\frac{15}{4}+\frac32\frac{r^2}{b^2}+\frac14\frac{r^2}{b^2}). \eqno(19)
$$
 Similarly from (12) and (2b) we find
$$
\bar h_{00}^{(2)}(m,r)=\frac{m^2G^2}{b^2}(\frac{15}{2}-6\frac{r^2}{b^2}+\frac{3}{2}\frac{r^4}{b^4}).                                                                                                       \eqno(12a)
$$
We note that $h^{(2)}_{\alpha\beta}(m,r)$ differs from harmonic  $ h^{(2)har}_{\alpha\beta}(m,r)$  and isotropic$h^{(2)iso}_{\alpha\beta}(m,r)$ only by gauge terms, in more detail see [1].
Equation (19) holds in all these systems.\\
{\bf Exterior solution.}
With the help of (5) from (10) we obtain the contribution to $\bar h^{(2)}_{\alpha\beta}(m',r) $ from $r'<b$
$$
16\pi G\int_{r'<b<r}\frac{d^3x'}{4\pi}\frac{1}{|\vec x-\vec x' |}(T^{(1)}_{\alpha\beta}+t_{\alpha\beta})=
(mG)^2[-\frac{14}{3}\frac{\delta_{\alpha\beta}}{rb}-\frac{4}{5\cdot7}b(\frac{x_{\alpha}x_{\beta}}{r^5}
-\frac{\delta_{\alpha\beta}}{3r^3})].                                                              \eqno(20)
$$
Here the equations (А4), (А5) and (А16) in [1] were used.
With the help of relation  (see for example equation  (70) in [5)])
$$
16\pi Gt_{\alpha\beta}=(mG)^2[\frac{14\delta_{\alpha\beta}}{r^4}-\frac{28x_{\alpha}x_{\beta}}{r^6}]\eqno(21)
$$
we find the contribution to $\bar h^{(2)}_{\alpha\beta}(m',r) $ from $ b<r'$
$$
16\pi G\int_{b<r',r}\frac{d^3x'}{4\pi}\frac{1}{|\vec x-\vec x' |}t_{\alpha\beta}(x')=
(mG)^2[\frac{14}{3}\frac{\delta_{\alpha\beta}}{rb}-\frac{7x_{\alpha}x_{\beta}}{r^4}+
Ъ\frac{28}{5}b(\frac{x_{\alpha}x_{\beta}}{r^5}
-\frac{\delta_{\alpha\beta}}{3r^3})].                                            \eqno(22)
$$
Here equations (A18) and (А15) in [1] were used.
The sum of these two contribution (20) and (22) gives
$$
\bar h^{(2)}_{\alpha\beta}(m',r)=(mG)^2[-7\frac{x_{\alpha}x_{\beta}}{r^4}+\frac{192}{35}b(\frac{x_{\alpha}x_{\beta}}{r^5}
-\frac{\delta_{\alpha\beta}}{3r^3})].                                            
.                     \eqno(23)
$$
We remind here that $\bar h^{(2)}_{\alpha\beta}(m',r)=\bar h^{(2)}_{\alpha\beta}(m,r)$
because $\bar h^{(2)}_{\alpha\beta}(m',r)=0$, see the second equation in (2). It is usefull to note that  the region $b<r'$ contributes to (23) essentially more then
$r'<b$. 

From the first equation in (13) and the relation
$$
h^{(2)}(m',r)=(mG)^2(\frac{12}{rb}+\frac{10}{r^2})=-\bar h(m',r)                   \eqno(24)
$$
it follows
$$
h^{(2)}_{\alpha\beta}(m',r)=m^2G^2[\frac{6\delta_{\alpha\beta}}{rb}+ (\frac{5\delta_{\alpha\beta}}{r^2}-
\frac{7x_{\alpha}x_{\beta}}{r^4})+\frac{192b}{35}(\frac{x_{\alpha}x_{\beta}}{r^5}-\frac{\delta_{\alpha\beta}}{3r^3})].
                                                                                                                \eqno(25)
$$
The expression for  $h^{(2)}_{\alpha\beta}(m,r)$ is obtained from here by dropping the term  $\frac{6\delta_{\alpha\beta}}{rb}$ because it is cancelled by term in $h^{(1)}_{\alpha\beta}(m',r)$:
$$
h^{(1)}_{\alpha\beta}(m',r)=h^{(1)}_{\alpha\beta}(m,r)-m^2G^2\frac{6\delta_{\alpha\beta}}{rb},
\quad h^{(1)}_{\alpha\beta}(m,r)=\delta_{\alpha\beta}\frac{2mG}{r}.                                  \eqno(26)
$$
Thus 
$$
h^{(2)}_{\alpha\beta}(m,r)=m^2G^2[(\frac{5\delta_{\alpha\beta}}{r^2}-
\frac{7x_{\alpha}x_{\beta}}{r^4})+\frac{64b}{35}(3\frac{x_{\alpha}x_{\beta}}{r^5}-
\frac{\delta_{\alpha\beta}}{r^3})].
                                                                                               \eqno(25a)
$$
It is easy to check that the interior solution (18) continuously goes over to exterior one  (25a):
$$
\left.h^{(2)}_{\alpha\beta}(m,r)\right|_{r\to b}=\left(\frac{mG}{b}\right)^2[\frac{111}{35}\delta_{\alpha\beta}-
\frac{53}{35}\frac{x_{\alpha}x_{\beta}}{b^2}]=\left.h^{(2)}_{\alpha\beta}(m,r)\right|_{b\gets r}.        \eqno(27)
$$

We note that the linear approximation (2) holds in harmonic, isotropic and considered here coordinates systems. Our system, given in  $G^2$ approximation by (18) and (25a) we will call the preferred one, because the gauge degrees of freedom do not contribute to it. 

Next we find the connection between radius of the matter ball  $a$ in standard system with radius $b$ in preferred system. The observable  (invariant) radius is given by the relation
$$
a_{obs}=a+\frac{mG}{3}+\frac{3}{10}\frac{m^2G^2}{a},                                              \eqno(28) 
$$
see equation (2.16) in [4] in $G^2$ approximation. Now we find it in our preferred system. Due to spherical symmetry it does not matter in what direction we measure the radius. We choose axis  1 for that direction. Then 
$$
dl=[1+h_{11}^{(1)}+h_{11}^{(2)}]^{1/2}dr=[1+\frac12h_{11}^{(1)}+\frac12h_{11}^{(2)}-\frac18h_{11}^{(1)2}]dr.
\eqno(29)
$$ 
Using (2) and (18), we find
$$
b_{obs}\equiv a_{obs}=\int_{0}^{b}dl=b+\frac{4}{3}mG+\frac{97}{70}\frac{m^2G^2}{b}.\eqno(29a)
$$
From (28) and (29a) we have
$$
a=b+mG+\frac{38}{35}\frac{m^2G^2}{b}.                                                              \eqno(30)
$$

It is interesting to find  outside the matter the connection between the coordinate  $r^{st}$ in standard system and coordinate $r$ in preferred system. To that end  we first find the expression for observable radius $r_{obs}$ in standard system
$$
r_{obs}=a_{obs}+\int_{a}^{r}dl=a_{obs}+\int_{a}^{r}dr(g_{rr}^{st})^{1/2}=a_{obs}+r-a+mG\ln\frac{r}{a}+
m^2G^2(\frac{3}{2a}-\frac{3}{2r}).                                                                    \eqno(31)
$$
Here everywhere  $r=r^{st}$, $ g_{rr}^{st}=(1+2\phi)^{-1}=1-2\phi+(2\phi)^2$, $\phi=-mG/r$. In preferred system we get 
$$
r_{obs}=r+mG(\frac43+\ln\frac{r}{b})+m^2G^2(\frac{4}{5b}+\frac{3}{2r}-\frac{32}{35}\frac{b}{r^2}). \eqno(32)
$$
Equating  (31) and (32), we get the relation  $r$ with $r^{st}$. To simplify this relation we note that up to terms of order $mG$ we have  $a=b+mG$, see (30). Then with this accuracy $r^{st}=r+mG$. So
$$
\ln\frac{r^{st}}{a}=\ln\frac{r}{b}+mG(\frac1r-\frac1b).                                            \eqno(33)
$$
Thus we get
$$
r^{st}=r+mG+m^2G^2(\frac2r -\frac{32}{35}\frac{b}{r^2}).                                    \eqno(34)
$$
Using this relation I failed to obtain the preferred system from the standard one. So these systems are different.

If the exterior solution  $h^{(1)}_{\alpha\beta}(m,r)$ satisfies Hilbert condition, then for the movement of nonrelativistic particle only  $h^{(2)}_{00}(m,r)=-2m^2G^2/r^2$ is essential.  In other words terms 
$h^{(2)}_{\alpha\beta}(m.r)$, ($\alpha.\beta=1,2,3.$) are essential for moving of relativistic particle, when 
 $v/c$ is not small.

 Now we note that from theory the observable length of a circle with radius $r$ is given by the expression
$$
L_{obs}(r)=2\pi[r+mG+m^2G^2(\frac{2}{r}-\frac{32}{35}\frac{b}{r^2})]. \eqno(35)
$$
From this relation in principle one can obtain the radius $b$ of matter ball.
The function  $L_{obs}(r)$ is obtainable from measurement. In general relativity it is impossible to get this radius
from measurement outside the ball, see   Birkhoff therem in \S 32.2 in [5]. Our consideration shows that in $G^2$ approximation besides Schwarzschild metric there is the preferred metric which also satisfies gravitational equation,

Now I indicate how to obtain equation (35). The starting relation is
$$
dl^2=g_{\alpha\beta}dx_{\alpha}dx_{beta}=[\delta_{\alpha\beta}+h^{(1)}_{\alpha\beta}+h^{(2)}_{\alpha\beta}+\cdots]
dx_{\alpha}dx_{\beta}.                                                      \eqno(36)
$$
On the circle
$$
x_1=r\cos\varphi, \quad x_2=r\sin\varphi, \quad x_3=0,\quad \delta_{\alpha\beta}dx_{\alpha}dx_{beta}=(r\varphi)^2                                                                                                     \eqno(37)
$$
the terms proportional  $x_{\alpha} x_{\beta}$ in $g_{\alpha\beta}$ do not contribute to  $dl$. Then, see (26) and 
(25a)
$$
dl^2=(rd\varphi)^2[1+\frac{2mG}{r}+m^2G^2(\frac{5}{r^2}-\frac{64}{35}\frac{b}{r^3})]. \eqno(38)
$$
From here with considered approximation we have
$$
dl=rd\varphi[1+\frac{mG}{r}+m^2G^2(\frac{2}{r^2}-\frac{32}{35}\frac{b}{r^3}].\eqno(39)
$$
Integration over $\varphi$ gives (35).

Another in principle possible way of verifying the theory is the following. Using next iteration we can get $h_{00}^{(3)}(m,r)$. In this approximation sources will depend on $b$. Now in relation
$\sqrt{|g_{00}|}=\omega_0/\omega$ the function on the left is known from theory and the one on the right from measurement of gravitational shift of frequency from far of source.


It is  interesting to note that from differential gravitational equation naturally follows the solution (25a),
in which  $b=0$. This slution is also formed only by sources and it satisfies the Hilbert condition. I obtained it early, see. equation  (73) in [6] or (28) in [7]. In order this exterior solution joint smoothly with the interior solution,
one ought to do in  (18) some gauge transformation. Then the interior solution will not satisfy Hilbert condition. On this ground also exterior solution has smaller chance to correspond  Nature. 

In conclusion I note that qualitatively the dependence of exterior metric on $b$ will remain if the 3-graviton vertex of general relativity is replaced by the one suggested by field theory approach [7].
\section*{References}
1. A.I.Nikishov. arXiv: 1605.06305 v.I [physics gen -ph] 16 May 2016.\\
2.J.Schwinger.  Particles, sources. and fields. Addison- Wesley, 1970.\\
3. S. Weinberg. Gravitation and Cosmology. New York 1972.\\
4. M.J. Duff. Phys.Rev. D, 7, 2317, (1973).\\
5. C.W. Misner, K.S. Thorn, J.A. Wheeler. Gravitation. San Francisco, 1973.\\
6. A.I.Nikishov. Part. Nucl. v.32, No  1, p.5 (2001).  \\
7. A.I.Nikishov. Part. Nucl., v.37, No 5, p.1466  (2006). \\
\end{document}